\documentclass[fleqn,showkeys,preprint]{revtex4}
\usepackage{amssymb}
\usepackage{amsmath}
\usepackage{bm,color}
\usepackage{multirow}
\usepackage{mathtools}
\def\BigRoman{\uppercase\expandafter{\romannumeral\number\count255}}
\def\Romannumeral{\afterassignment\BigRoman\count255=}

\newcommand{\abs}[1]{\left\lvert #1 \right\rvert}
\newcommand\underrel[2]{\mathrel{\mathop{#2}\limits_{#1}}}

\begin{document}
\setcounter{page}{1}
\title[]{A mini-review of the diffusion dynamics of DNA-binding proteins: Experiments and models}
\author{Seongyu Park}
\affiliation{Department of Physics, Pohang University of Science and Technology (POSTECH), Pohang 37673, Republic of Korea}
\author{O-chul Lee}
\affiliation{Department of Physics, Pohang University of Science and Technology (POSTECH), Pohang 37673, Republic of Korea}
\author{Xavier Durang}
\affiliation{Department of Physics, Pohang University of Science and Technology (POSTECH), Pohang 37673, Republic of Korea}
\author{Jae-Hyung Jeon}
\email{jeonjh@postech.ac.kr}
\affiliation{Department of Physics, Pohang University of Science and Technology (POSTECH), Pohang 37673, Republic of Korea}
\date[]{}

\begin{abstract}
In the course of various biological processes, specific DNA-binding proteins must find a particular target sequence/protein or a damaged site on the DNA efficiently. DNA-binding proteins perform this task based on diffusion. Yet, investigations over recent decades have found that the diffusion dynamics of DNA-binding proteins are generally complicated and, further, protein-specific. In this review, we collect the experimental and theoretical studies that quantify the diffusion dynamics of DNA-binding proteins and review them from the viewpoint of diffusion processes.
\end{abstract}



\maketitle

\section{INTRODUCTION}
The biological processes that occur inside a nucleus, such as DNA replication, damage repair, and regulating gene expression, are initiated from the binding of specific DNA-binding proteins onto a specific DNA site. How can a protein find its target DNA sequences amid the vast number of non-target sites along the DNA? This problem has been explored for several decades, but the quantitative understanding of the whole target-search process remains elusive.  

In 1970, Riggs \textit{et al.}\cite{riggs1970} performed pioneering experiments on the target search dynamics of a Lac repressor, which led to the birth of the \textit{Facilitated Diffusion} model---the key idea explaining the DNA target-search process. They found that the association rate of a Lac repressor with the target site is 100-times faster than that of a 3D diffusion-limited search in the solution expected from the Smoluchowsky theory~\cite{riggs1970, berg1981}. To explain the experimental observation, Berg, Winter, and von Hippel introduced the notion of \textit{Facilitated Diffusion}~\cite{berg1981}. Simply put, the main idea of this model is that a DNA-binding protein finds its target by a combination of 3D diffusion in the solution and 1D diffusion along the DNA contour~\cite{riggs1970, berg1981, halford2004, kolomeisky2011}. The search time can be significantly reduced compared to the Smoluchowsky time since the protein can substantially increase the probability of hitting the target via the one-dimensional diffusion along the DNA if the target site is not far from the site where the protein locates. The Facilitated Diffusion model will be explained further in the following section.

In 1993, Kabata \textit{et al.}, for the first time, visualized a 1D translocation of RNA polymerase holoenzyme whose transcription activity is blocked. This experiment verified the 1D diffusion dynamics of a DNA-binding protein that the Facilitated Diffusion model is based on. Beyond this observation, the development of the single-molecule experimental technique made it possible to pursue further unexplored, relevant questions such as: how a protein can recognize the correct target sequences or whether a protein is continuously in contact with the DNA during diffusion or hopping to overcome the roadblocks on the DNA. 
Some theoretical studies have pointed out the possibility that the protein will feel a sequence-dependent DNA--protein interaction to identify the target sequence and that it could have multiple dynamic modes to improve the efficiency of the target search~\cite{barbi2004, barbi2004model, slutsky2004, mirny2009}. 
To date, there have been numerous experimental studies exploring the diffusion patterns and mechanisms of various DNA-binding proteins based on the single-molecule experimental tools. These tools include, e.g., total internal reflection fluorescence Microscopy~(TIRF), oblique-angle fluorescence imaging, and F\"{o}rster resonance energy transfer~(FRET), which typically probe the motion of a single protein with a spatial resolution of $\sim10~\mathrm{nm}$ and time resolution of $\sim50~\mathrm{ms}$. Through such experiments, the protein diffusion has been investigated at multiple timescales, from an order of $10~\mathrm{ms}$ to tens or hundreds of seconds, or even to an order of $100~\mathrm{min}$~\cite{kabata1993, gorman2008, tafvizi2008, tafvizi2011p53, leith2012, murata2015, itoh2016, murata2017, subekti2017, kong2016, cheon2019, graneli2006, kim2007, bonnet2008, biebricher2009, dikic2012, dunn2011, nelson2014, blainey2006, vestergaard2018, wang2006, marklund2020, gorman2007, gorman2010, brown2016, kamagata2018, wang2013, lin2014, lin2016, cho2012, jeong2011, kad2010, ghodke2014, hughes2013, liu2017b, cuculis2016, cuculis2015}.
These experimental investigations have demonstrated that the one-dimensional diffusion patterns of DNA-binding proteins are typically more complex than the simple Brownian motion initially conjectured in the Facilitated Diffusion model. The complexity of the motion of the protein seems to arise from various underlying heterogeneities, such as the sequence-dependent interactions between the protein and the DNA, conformational changes of the protein, the roadblock molecules on the DNA, or the bending of the DNA.
It is also an interesting finding from the experimental studies so far that the diffusion dynamics are highly protein-specific, presumably because of its distinct molecular complexity and biological role.

In this review, our aim is to collect information on the 1D diffusion dynamics of the DNA-binding proteins that have been experimentally studied and to provide an overview of the complex diffusion dynamics of the DNA-binding proteins (see Table~1).
It should be emphasized that our focus is not to review the experiments investigating the validity of the Facilitated Diffusion model or the target-search process itself, which are already available in the literature. This review focuses on the diffusion dynamics of a DNA-binding protein itself from the viewpoint of stochastic processes. Accordingly, in this review, we additionally introduce a number of mathematical diffusion processes that are helpful in describing the real diffusion processes of the DNA-binding proteins that have been experimentally observed. Also, we provide a mini-review of some theoretical models that are related to or aimed at the modeling of the complex diffusion dynamics of the proteins found in the experiments.  

This paper is organized in the following order. In Section~II, we introduce the four distinct diffusion modes in the Facilitated Diffusion model. Each diffusion mode is explained in terms of physical concepts and the protein groups that will appear throughout the paper. In Section~III, we provide a brief review of several diffusion models introduced in physics and mathematics for explaining normal and anomalous diffusion processes. The theoretical concepts and terminology introduced in this section will be used in the subsequent sections on the diffusion dynamics of DNA-binding proteins. In Section~IV, we review the diffusion properties of the DNA-binding proteins that have been reported to perform an ordinary Fickian (i.e., Brownian) motion along the DNA. The proteins are classified and explained according to the protein group. In Section~V, we deal with the proteins whose diffusion dynamics have been found to be non-Fickian.  The details of the observed anomalous dynamics are described according to the protein group. In the last main section of this paper (Section~VI), we review a series of theoretical diffusion models and their properties which seem relevant for explaining or understanding the anomalous diffusion dynamics of the DNA-binding proteins at various levels. Finally, a summary of this review is given in Section~VII. 


\section{Diffusion modes of DNA-binding proteins}
\label{sec:diffusionmodes}
\begin{figure}
\centering
\includegraphics[width=0.5\textwidth]{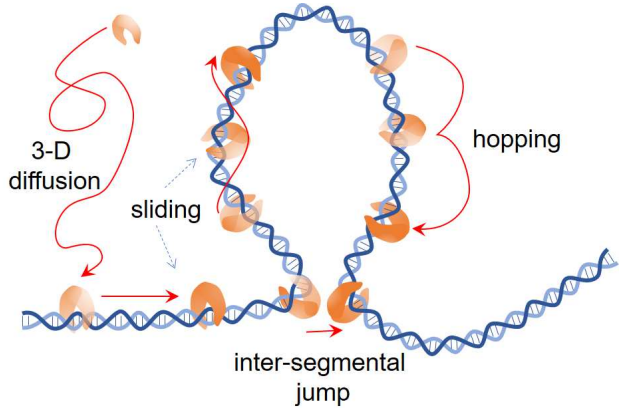}
\caption{The four diffusion modes of a DNA-binding protein. In the Facilitated Diffusion model, it is suggested that the DNA-binding proteins search on a DNA by sliding (including rotation-coupled sliding), hopping, and inter-segmental jumps. After dissociation from the DNA, the protein diffuses in the solution and can bind onto the DNA again and repeat the 1D search until it finds the target sequence. }
\label{fig1}
\end{figure}

As evidenced by the pioneering work on the search kinetics of DNA-binding proteins (such as the Lac repressor) for finding their target sites~\cite{riggs1970}, the experimentally measured search times are shorter than the value predicted based on a diffusion-limited search in three-dimensional space, by a factor of 10--100. This suggests that there are several possible routes for the DNA-binding protein to find its specific target site, and the search time can be significantly reduced by suitably combining these routes. This idea was posited in the Facilitated Diffusion model~\cite{richter1974, berg1976, berg1981}, which states that the DNA-binding protein typically finds its target based on a combination of a one-dimensional search along the DNA and a three-dimensional excursion (Fig.~\ref{fig1}). A simplified picture of this model is as follows: a target search protein freely diffuses in the bulk solution until it binds onto the DNA. Once it binds to a nonspecific site on the DNA, it randomly moves along the DNA until it dissociates from it, and repeats this cycle until it eventually finds the target site. In this search process, it seems that there are four distinct types of search strategy. We will briefly introduce these four diffusion modes.

\subsection{One-dimensional sliding along DNA}
A DNA-binding protein is able to associate with a DNA through a non-specific binding, mainly, due to the electrostatic interactions between the DNA phosphates and the basic residues of the protein~\cite{berg1981, barbi2004model, von2004}. In such a state, the protein can diffuse along the DNA in constant contact with the DNA phosphate, which is called the one-dimensional sliding motion of the protein. Fig.~\ref{fig1} shows a schematic of this process. One-dimensional sliding is the essential part in Facilitated Diffusion: it significantly increases the probability of finding the target site compared to the 3D-limited diffusion, by allowing the protein to repeatedly sample the DNA sequences around the protein. This process was observed for the first time in 1993 at the single-molecule level with an RNA polymerase translocating along DNA~\cite{kabata1993}.

Depending on the protein architecture and the way it associates with the DNA, the sliding diffusion can be linear or helical (see Fig.~\ref{fig1}). Linear sliding is when the protein moves along the DNA in a bi-directional way. Helical sliding is a rotation-coupled linear diffusion such that the protein rotates around the DNA groove while moving along the DNA. In this case, the diffusion constant is expected to scale as $D\sim 1/R^3$ with the  radius $R$ of the protein, not as $D\sim 1/R$ by the Stokes--Einstein law for simple diffusion~\cite{bagchi2008}. The diffusion constant for 1D rotation-coupled diffusion can be described as in~\cite{blainey2009}:

\begin{figure}
\centering
\includegraphics[width=0.4\textwidth]{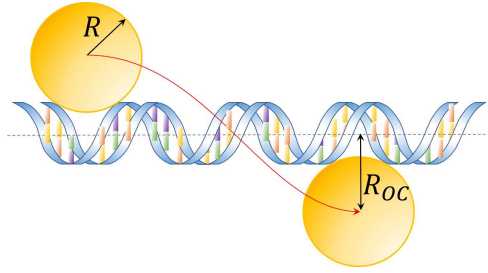}
\caption{Helical sliding mode of a DNA-binding protein. $R$ is the radius of the DNA binding protein. $R_\mathrm{OC}$ is the distance between the center of mass of the protein and the DNA axis.}\label{fig2}
\end{figure}

\begin{eqnarray}
\label{eqn:helical}
D\approx b^{2}\frac{k_{B}T}{8\pi\eta R^{3}+6\pi\eta R(R_{\mathrm{OC}})^{2}}\mathrm{exp}[-(\epsilon/k_{B}T)^{2}],
\end{eqnarray}
where $R_{\mathrm{OC}}$ is the minimum distance from the protein's center of mass to the DNA axis, $\eta$ is the viscosity of water, and $b$ is the distance per full rotation~($\sim3.4~\mathrm{nm}$), see Fig~\ref{fig2}. In this formula, the non-Arrhenius factor $\mathrm{exp}[-(\epsilon/k_{B}T)^{2}]$ is due to the rugged Gaussian random potential, where $\epsilon$ is the root mean square variation in the protein--DNA interaction energies~\cite{zwanzig1988, goychuk2014, goychuk2018, banerjee2014}. For further information, see Sections~III and VI. 

It was inferred that the diffusion constants of some proteins (e.g., hOGG1, LacI, Fpg, Nei, Nth, EcoRV, TRF1, TRF2) have the dependence $1/R^{3}$~\cite{blainey2009, dunn2011, nelson2014, dikic2012, lin2014}, which implies that these proteins move along the DNA using the rotation-coupled sliding mode~\cite{blainey2009}. A computational study by Terakawa \textit{et al.} reported that a tumor suppressor p53 performs a rotational diffusion at low-salt concentrations while the diffusion is no longer helical at high-salt concentrations~\cite{terakawa2012}. In \cite{cho2012, jeong2011}, it was found that the MutS protein shows rotation-coupled sliding based on a single-molecule study using a single-molecule fluorescence polarization total internal reflection fluorescence (smPolarization-TIRF) and a fluorescence resonance energy transfer (FRET).

As will be explained in detail below, sliding diffusion can be either sequence-independent or sequence-dependent, depending on the ambient conditions as well as on the structure of the protein~\cite{wang2011}. Sequence-independent sliding can be understood as an unbiased random walk or Brownian motion. Then, the sliding dynamics is a Gaussian process, with the mean-squared displacement (MSD) increasing linearly with time $t$, i.e., $\langle  x^{2}(t) \rangle =2D t$~\cite{halford2004}. When the sliding dynamics is sequence-dependent, on the other hand, the sliding dynamics is  subdiffusive~\cite{barbi2004, barbi2004model, goychuk2014, goychuk2018} at certain time scales, where the MSD grows with $t$ as $\langle  x^{2}( t)\rangle \propto t^{\alpha}$ with the exponent $\alpha \in (0, 1)$~\cite{barbi2004, barbi2004model, wang2011, sinai1983}.

\subsection{Hopping along DNA}
The hopping mechanism is another 1-D diffusion mechanism, distinct from sliding. As illustrated in Fig.~1, in this mechanism, a protein moves along the DNA through repeatedly dissociating and associating with the DNA. However, it is not easy to tell whether the 1D motion is sliding or hopping because of the limitations of the temporal and spatial resolutions in the experiment~\cite{wang2011, ober2015}. Thus, indirect methods are often employed to examine whether a protein is indeed frequently dissociated from and associated to the DNA during its 1D diffusion. In these experiments, obstacles, such as nucleosomes or EcoRI, are placed along the DNA~\cite{hedglin2010, gorman2010, pluciennik2007}. If a protein diffusion is mainly accomplished via the hopping mechanism, it can easily jump over the obstacles and thus obtain a diffusion constant insensitive to the obstacle density. 
The dependence of $D$ on the salt concentration is another signature test for hopping diffusion~\cite{cheon2019, tafvizi2011p53, kong2016}. Since the stability of the protein--DNA nonspecific complex is mainly due to electrostatic interactions~\cite{halford2009, hedglin2010, barbi2004model, barbi2004}, an increase in the salt concentration is expected to increase the dissociation rate of the protein so that it spends more time detached from the DNA. The diffusion constant increases accordingly if the protein diffusion is governed by the hopping mechanism.

There are a few proteins that seem to show hopping diffusion. Mlh1-Pms1, a ring-shaped mismatch repair protein, has a diffusion constant that depends on the salt concentration, and freely bypasses obstacles, while other proteins, such as Msh2-Msh6, which wraps around the DNA and intimately contacts with the phosphate backbone, show rare bypassing events over the obstacles~\cite{gorman2010}. Glycosylases such as human alkyladenine DNA glycosylase (AAG) and uracil DNA glycosylase (UDG) are known to sample only one nucleotide of a base pair at a time. If the protein slides along the phosphate backbone, it can search only one strand of the DNA. Nevertheless, they can find a lesion on both sides of the double strands~\cite{hedglin2010, porecha2008}, which implies that they can hop during the 1-D target searching process.

\subsection{Inter-segmental transfer}

Inside a cell, the DNA exists in the form of a globule, where the one-dimensional DNA has many loops and allows physical contacts between distant segments. During the one-dimensional diffusion along the DNA, a DNA-binding protein can make a short hop around one of these contact points so that it jumps from one place to a considerably distant place along the DNA contour. Fig.~\ref{fig1} depicts such an inter-segmental jump of a DNA-binding protein. This process allows the DNA-binding protein to have intermittent long jumps like L\'{e}vy flights during its one-dimensional search~\cite{lomholt2009}.

\subsection{Three-dimensional diffusion in bulk solution}
The proteins in a bulk solution can find their target sites directly by a three-dimensional diffusion (Fig.~1). It is usually supposed to be a Brownian motion where the particle diffuses in the solution without making any interactions with the DNA strand.
The diffusion-limited association rate is then given by
\begin{equation}
k=4\pi D_{3d} ba
\end{equation}
where $D_{3d}$ ($\approx~10^{-10}~\mathrm{m}^{2}/\mathrm{s}$ for a protein with a diameter of $5~\mathrm{nm}$) is the diffusion constant of a protein in the bulk, $b$ is the size of the target site ($\approx0.34$~nm for a single nucleotide), and $a$ is the fraction of the reactive surface of the protein ($\approx 0.2$--$0.5$)~\cite{smoluchowski1918, mirny2009}.
For the given constants, the association rate is estimated to be $k\approx 10^{8}~ \mathrm{M}^{-1} s^{-1}$, which is $10$--$100$ times smaller than the value $k_{\mathrm{exp}}\approx 10^{10}~\mathrm{M}^{-1}s^{-1}$ observed by Riggs \textit{et al.}~\cite{riggs1970}. As one can see from this discrepancy, diffusion-limited association is not the way for a protein to find its target sites. The 3D diffusion of a protein must be followed by a 1D diffusion through the non-specific binding to the DNA~\cite{berg1981}.


\section{ Diffusion processes and mathematical models }
\label{sec:background}
In this section, we provide a brief overview of the diffusion processes introduced in the literature that are relevant to quantifying the diffusion dynamics of DNA-binding proteins observed in the experiments. In order to quantify a diffusion, a standard measure is the mean-squared displacement (MSD). On many occasions, the MSD growth scales with time as
\begin{equation}
\label{eq:anomalousmsd}
\langle x^2(t)\rangle =2D_\alpha t^\alpha.
\end{equation}
Here, $\alpha$ is called the anomaly exponent, and $D_\alpha$ is referred to as (generalized) diffusion constant of physical dimension $[\mathrm{m}^2/\mathrm{s}^\alpha]$. A diffusion process can be classified in terms of $\alpha$~\cite{haus1987, bouchaud1990, metzler2000}. It is called normal or Fickian when $\alpha=1$ (where $D_1\equiv D$ is, in particular, called the diffusion constant (in units of $[\mathrm{m}^2/\mathrm{s}]$) and is anomalous when $\alpha\neq1$. In particular, the case of $0<\alpha<1$ is called subdiffusion whereas the case of $\alpha>1$ is superdiffusion. As a particular limit, the dynamics with $\alpha=2$ is the ballistic motion where a particle moves at a constant velocity. Apart from this classification, recently there has been interest in whether a diffusive motion is homogeneous or spatiotemporally heterogeneous~\cite{cherstvy2013, wei2020, cherstvy2015, cherstvy2013b, chechkin2017, sposini2018, leibovich2019, cherstvy2014, salgado2016}. In soft complex systems, it has been found that some processes show the Fickian but non-Gaussian diffusion originating from spatiotemporal heterogeneity~\cite{wang2009, lampo2017, cherstvy2019, wang2012, hapca2009, chaudhuri2007, roldan2017, toyota2011, samanta2016}. 

\subsection{Gaussian Fickian diffusion ($\alpha=1$)}

This is the classical model explaining the Brownian motion of a colloid in water. Its dynamics can be described by the overdamped Langevin equation
\begin{equation}\label{eq:ole}
\gamma\frac{d}{dt}x(t)=\sqrt{2\gamma k_B T}\xi(t)    
\end{equation}
where $\xi$ is a Gaussian white noise that satisfies $\langle \xi \rangle=0$ \& $\langle \xi(t)\xi(t') \rangle=\delta(t-t')$, $\gamma$ the frictional coefficient of the particle, and $T$ the absolute temperature (with the Boltzmann constant $k_B$).  Then, the formal solution of this equation is
\begin{equation}\label{eq:ole2}
x(t)=\sqrt{\frac{2k_BT}{\gamma}}\int_0^t\xi(t')dt'+x_0.
\end{equation}
For simplicity, we set $x_0=0$ throughout this paper. The MSD is given by
\begin{equation}
\langle x^2(t)\rangle = \frac{2k_BT}{\gamma}t=2Dt
\end{equation}
where $D=k_BT/\gamma$ is the diffusivity (diffusion constant) given by the Einstein relation.  Since $\xi$ is a Gaussian process, $x(t)$ is also Gaussian. Using the variance relation above, the probability density function is easily identified to
\begin{equation}\label{eq:Pgauss}
P_G(x,t)=\sqrt{\frac{1}{4\pi D t}}\exp(-x^2/[4Dt]).
\end{equation}
It can be also shown that Eq.~\eqref{eq:Pgauss} is the solution of Fick's diffusion equation
\begin{equation}
\frac{\partial}{\partial t}P_G(x,t)=D\frac{\partial^2 }{\partial x^2}P_G(x,t)
\end{equation}with the initial condition $P(x,t\to 0)=\delta(x)$. 
Since $\xi$ is a $\delta$-correlated noise, $x(t)$ is a Markovian process having no memory in spatial increments. The covariance function is given by
\begin{equation}
\langle x(t)x(t')\rangle= x^2(\mathrm{min}(t,t'))
\end{equation}
and its velocity autocorrelation is $\langle \dot{x}(t)\dot{x}(t')\rangle \propto \delta(t-t')$.

\subsection{Gaussian non-Fickian diffusion ($\alpha\neq 1$) (viscoelastic subdiffusion)}
\label{sec:FLE}

As a generalized model for the Gaussian Fickian model above, one can think of a Gaussian, stationary increment process that has variance $\langle x^2(t)\rangle \sim t^{\alpha}$ with $\alpha\neq1$. One of the models in this class is viscoelastic subdiffusion, also known as fractional Brownian motion~\cite{mandelbrot1968}. Physically, this process can be understood as the motion of a Brownian particle in a viscoelastic environment. In this situation, the movement of a particle affects its environment, which in turn gives rise to a self-memory of the motion mediated by the environment. It is known that this model is adequate for the description of a biological molecule in a crowded or polymeric environment like a cytoplasm~\cite{metzler2014}.

In the overdamped limit, the fractional Langevin equation generalizing Eq.~\eqref{eq:ole} can be written in this form~\cite{metzler2014}:
\begin{equation}
\gamma_{\alpha'}\int^{t}_{0}(t-t')^{\alpha'-2} \left(\frac{dx(t')}{dt'}\right)dt'=\sqrt{\frac{\gamma_{\alpha'} k_BT}{\alpha'\left(\alpha'-1\right) D_{\alpha'}}} \xi_{\alpha'}(t).
\end{equation}
where $\gamma_{\alpha'}$ is a generalized frictional coefficient and $D_{\alpha'}$ is the generalized diffusivity defined above.  Here, $\xi_{\alpha'}$ is a fractional Gaussian noise of index $\alpha'$ defined in the interval $(1,~2)$. The amplitude in front of the noise is exactly given in order to satisfy the Kubo generalized fluctuation dissipation relation. As a result, the autocorrelation of $\xi_{\alpha'}$ satisfies the following relation [$1<\alpha'<2$]: 
\begin{equation}\label{eq:fgn}
\langle {\xi}_{\alpha'}(t_1){\xi}_{\alpha'}(t_2) \rangle = \alpha' \left(\alpha'-1\right)D_{\alpha'} {\left|t_{1}-t_{2}\right|} ^{\alpha'-2}
\end{equation}
for $t_1\neq t_2$. In the limit of $\alpha'=1$, $\langle {\xi}_{\alpha'}(t_1){\xi}_{\alpha'}(t_2) \rangle =2D_{\alpha'}\delta(t_1-t_2)$. 
The above fractional differential equation can be solved (see \cite{metzler2014, jeon2012, jeon2010} for the derivation). From this solution, one can obtain that the MSD of such a process is expressed in terms of the generalized Mittag-Leffler function, which scales as
\begin{equation}
    \langle x^2(t) \rangle \sim t^{\alpha}~\hbox{with}~\alpha=2-\alpha'.
\end{equation}
Thus, it shows that the persistent fractional Gaussian noise $\xi_{\alpha'}$ with $1<\alpha'<2$ leads to subdiffusion with the anomaly exponent $\alpha$ above. 

\subsection{Heterogeneous diffusion}
\label{sec:nonGaussian}
In biological systems, diffusion is often not only anomalous but also spatiotemporally heterogeneous~\cite{metzler2014, sposini2018, wang2009, wang2012}. Below we introduce three diffusion models dealing with different heterogeneities in the diffusion process.

\subsubsection{Generalized grey Brownian motion} 
Generalized grey Brownian motion is a process defined by the stochastic equation
\begin{equation}
x(t)=\sqrt{2D_{i}}\int_{0}^{t}\xi(t')dt'
\end{equation}
with a random diffusion constant $D_i$, given by a probability function $p_D(D)$ such as an exponential or generalized Gamma distribution~\cite{sposini2018}. Seemingly similar to the ordinary Brownian (or overdamped Langevin) motion Eq.~\eqref{eq:ole2}, due to the random diffusion constant, this process has different statistical features from the former homogeneous process. In physical systems, this model describes the statistics for an ensemble of ordinary Brownian walkers with a spectrum of  diffusion constants or the case that identical particles are placed in locally different environments. The MSD of this process is given simply by
\begin{equation}\label{eq:msdDD}
\langle x^2(t) \rangle =2t\left[\int_0^\infty D' p_D(D') dD'\right]=2\langle D \rangle_\mathrm{st} t.
\end{equation}
Thus, on average, the process appears to be a Fickian dynamics with the average diffusion constant $\langle D \rangle_\mathrm{st}$. However, the process is not Gaussian in general. The propagator $P(x, t)$ for an ensemble of such processes is given in the super-statistical sense as~\cite {beck2001, beck2003, sposini2018}:
\begin{equation}
\label{eqn:superstatistic}
    P(x, t)=\int_{0}^{\infty}p_{D}(D') P_G(x, t|D') dD',
\end{equation}
where $P_G$ is the Gaussian PDF defined in Eq.~\eqref{eq:Pgauss}. A well-known example is the case of an exponentially decaying $p_D$~\cite{wang2009, wang2012, chechkin2017, sposini2018}. In this case, $P(x,t)$ is given to be exponential, which is known as a Laplace distribution.

\subsubsection{Fluctuating diffusivity model}
\label{sec:fluctuatingdiffusivity}
While generalized grey Brownian motion deals with particle-to-particle variation or spatial heterogeneity in the diffusion process, fluctuating diffusivity model incorporates the temporal heterogeneity of a diffusion process. This process is defined through the stochastic equation
\begin{equation}
\label{eqn:DDlangevin}
    \frac{d}{dt}x(t)=\sqrt{2D(t)}\xi(t),
\end{equation}
where the diffusion constant is now a time-dependent stochastic variable. The temporal fluctuation in $D(t)$ is modeled by a stochastic process that is always positive definite and has a stationary state. For example, in the minimal model~\cite{chubynsky2014, chechkin2017}, it is modeled by $D(t)=Y(t)^2$ where $Y(t)$ is an Ornstein--Uhlenbeck process. Defining $\tau_c$ as the cross-over time after which $D(t)$ reaches a stationary state with the stationary distribution $p_D(D)$, it turns out that the fluctuating diffusivity model behaves differently for short times $t<\tau_c$ and long times $t>\tau_c$. (1) Short-time: if $D(t)$ starts at $t=0$ with a stationary initial condition, the propagator $P(x,t)$ becomes identical to that of generalized grey Brownian motion, Eq.~\eqref{eqn:superstatistic}. Therefore, at this time scale, the process is non-Gaussian. (2) Long-time: $D(t)$ is in a stationary state with the average diffusivity $\langle D \rangle_\mathrm{st}$. The process then becomes Gaussian with the propagator  
\begin{equation}
P(x,t)=\sqrt{\frac{1}{4\pi \langle D \rangle_\mathrm{st} t}}\exp(-x^2/[4\langle D \rangle_\mathrm{st}t]).
\end{equation}
At all times, the MSD exhibits Fickian scaling, precisely given by Eq.~\eqref{eq:msdDD}.

\subsubsection{The Sinai diffusion model}
A heterogeneous diffusion can occur in a disordered medium where the local environmental state is randomly given but quenched. A discrete version in this class of models is the Sinai diffusion model introduced in 1983~\cite{sinai1983}. Consider a random walk on a one-dimensional disordered lattice where each site is governed by a randomly chosen bias field $\eta_i$, with $-1\leq \eta_{i} \leq 1$. 
At site $i$, the walker jumps to its neighboring sites $i+1$ or $i-1$ with the transition rate
\begin{align}
\begin{split}
    \omega(i\to i+1)=\frac{1+\eta_{i}}{2},\\
    \omega(i\to i-1)=\frac{1-\eta_{i}}{2}.
\end{split}
\end{align}
Accordingly, for a given random field, a site-specific asymmetric net current is induced. In the continuum limit, the corresponding diffusion process is given by the Fokker--Planck equation
\begin{equation}
\frac{\partial}{\partial t}P(x,t)=-\frac{\partial }{\partial x}[\eta(x)P]+\frac{1}{2}\frac{\partial^2}{\partial x^2}P.
\end{equation}
In this picture, the Sinai model can be interpreted as a diffusion in a random potential $U(x)=-\int_{-\infty}^x dx' \eta(x')$. 

Now imagine a set of random fields where the bias field $\eta_i$ is given by a symmetric distribution $P(\eta)$ satisfying $\int_{-1}^1 d\eta P(\eta) \mathrm{ln}\left(\frac{1+\eta}{1-\eta}\right)=0$. It has been demonstrated that the MSD in the long-time limit behaves as~\cite{sinai1983, blumberg1989, kesten1986, goychuk2017, metzler2014}.
\begin{equation}\label{eq:sinai}
\overline{\langle x^{2}(t)\rangle}\sim (\mathrm{ln}~t)^{4}.
\end{equation}(The overline over the MSD indicates the disorder-average over the sample). This logarithmic time dependence can be physically understood in the context of diffusion in a random potential. For an interval of length $\Delta x$, the expected time to escape from this interval scales as $t(\Delta x)\sim \exp(-\Delta U)$ where $\Delta U=U(x_0+\Delta x)-U(x_0)$ is the potential difference. Since the random potential is given symmetrically, $\Delta U$ on average increases as $(\Delta x)^{1/2}$. Therefore, the logarithmically increasing diffusion Eq.~\eqref{eq:sinai} is obtained.

\begin{table*}
\label{table}
\caption{Classification of diffusion dynamics and modes of DNA-binding proteins. In column 3: the diffusion dynamics of a protein is classified into Fickian ($\alpha=1$), non-Fickian ($\alpha\neq 1$), and mixed (having both Fickian and non-Fickian characteristics). In column 4: the diffusion modes of a protein are classified into single (only one mode exists) and multiple (several modes exist or a single mode has multiple dynamic states). In column 6: the possibility of rotation-coupled sliding is indicated. Here, the mark $\bigcirc$ means that rotational sliding was observed in the experiment, or conjectured. In the table, a blank element refers to there being unidentified or unclear features from the references cited.}
{\scriptsize
\begin{tabular*}{\textwidth}{l@{\extracolsep{\fill}} c c c c c c c c}
\multicolumn{1}{c}{group} & \multicolumn{1}{c}{protein} & \multicolumn{1}{c}{\begin{tabular}[c]{@{}l@{}}Fickian/\\ non-Fickian\end{tabular}} & \multicolumn{1}{c}{\begin{tabular}[c]{@{}c@{}}diffusion \\ mode\end{tabular}} & \multicolumn{1}{c}{\begin{tabular}[c]{@{}c@{}}dominant\\ mode\end{tabular}} & \multicolumn{1}{c}{\begin{tabular}[c]{@{}c@{}}rotation-\\coupled\\ sliding\end{tabular}} & \multicolumn{1}{c}{\begin{tabular}[c]{@{}c@{}}observation\\ time window\end{tabular}} & \multicolumn{1}{c}{Ref.} \\
\hline\hline 
\multicolumn{1}{c}{\multirow{3}*{p53}} & WT & Fickian & multiple & sliding &  & 33~ms $\sim$ 5~s & \cite{tafvizi2008, murata2015, tafvizi2011p53,terakawa2012,kamagata2017}\\  
 & TC domain & Fickian & single & sliding &  & 33~ms $\sim$ 5~s & \cite{murata2015,murata2017,tafvizi2011p53,terakawa2012}\\
 & NCT domain & Fickian & multiple & hopping &  & 33~ms $\sim$ 5~s & \cite{murata2017,tafvizi2011p53,terakawa2012,kamagata2017}\\
\hline
\multicolumn{1}{c}{\multirow{3}*{\begin{tabular}[c]{@{}c@{}c@{}}Architectural\\DNA-binding\\proteins\end{tabular}}} & Nhp6A & Fickian & single & sliding & $\bigcirc$ & 70~ms $\sim$ 1.6~s & \cite{kamagata2018}\\
 & HU & Fickian & single & sliding & $\bigcirc$ & 70~ms $\sim$ 1.6~s & \cite{kamagata2018, tan2016}\\
 & Fis & Fickian & multiple & sliding & $\bigcirc$ & 70~ms $\sim$ 1.6~s & \cite{kamagata2018}\\
\hline
\multicolumn{1}{c}{\multirow{5}*{\begin{tabular}[c]{@{}c@{}c@{}}Mismatch\\repair\\proteins\end{tabular}}}  & MutS with ADP & Fickian & single & sliding & $\bigcirc$ & 50~ms $\sim$ 10~min & \cite{cho2012, jeong2011}\\
 & MutS with ATP & Fickian & single & hopping &  & 50~ms $\sim$ 10~min & \cite{cho2012}\\
 & Msh2-Msh6 & Fickian & single & sliding &  & 0.2~s $\sim$ 10~min & \cite{gorman2007, gorman2010, brown2016}\\
 & Msh2-Msh3 & Fickian & multiple & mixed &  & 0.2~s $\sim$ 10~min & \cite{brown2016}\\
 & Mlh1-Pms1 & Fickian & & hopping &  & 0.2~s $\sim$ 7~min & \cite{gorman2010}\\
\hline
\multicolumn{1}{c}{\multirow{5}*{glycosylases}}  & hOGG1 & Fickian & multiple & sliding & $\bigcirc$ & 10~ms $\sim$ 5~s &\cite{blainey2009, hedglin2010, blainey2006,vestergaard2018}\\
 & Fpg & mixed & multiple & sliding & $\bigcirc$ & 15~ms $\sim$ 1~min & \cite{dunn2011,nelson2014}\\
 & Nei & mixed & multiple & sliding & $\bigcirc$ & 15~ms $\sim$ 1~min & \cite{dunn2011,nelson2014}\\
 & Nth & mixed & multiple & sliding & $\bigcirc$ & 15~ms $\sim$ 1~min & \cite{dunn2011,nelson2014}\\
\hline
\multicolumn{1}{c}{\begin{tabular}[c]{@{}c@{}c@{}}restriction\\enzyme\end{tabular}}  & EcoRV & Fickian & multiple & sliding & $\bigcirc$ & 20~ms $\sim$ 30~s & \cite{bonnet2008,biebricher2009,dikic2012}\\
\hline
\multicolumn{1}{c}{\begin{tabular}[c]{@{}c@{}c@{}}Lac\\repressor\end{tabular}}  & LacI & Fickian & multiple & mixed & $\bigcirc$ & 10~ms $\sim$ 10~min & \cite{blainey2009, wang2006, marklund2020, hammar2012}\\
\hline
\multicolumn{1}{c}{RecA}  & Rad51 & Fickian &  &  &  & 124~ms $\sim$ 12.4~s & \cite{graneli2006}\\
\hline
\multicolumn{1}{c}{\multirow{2}*{\begin{tabular}[c]{@{}c@{}c@{}}XPC\\homologs\end{tabular}}}  & rad4-rad23 & mixed & multiple & mixed &  & 80~ms $\sim$ 5~min & \cite{kong2016}\\
 & XPC-rad23 & mixed & multiple & mixed &  & 0.1~s $\sim$ 5~min & \cite{cheon2019}\\
\hline
\multicolumn{1}{c}{\multirow{3}*{\begin{tabular}[c]{@{}c@{}c@{}}Telomeric sequence\\ binding proteins\end{tabular}}}  & TRF1 & non-Fickian & multiple & sliding & $\bigcirc$ & 0.5~s $\sim$ 100~s & \cite{lin2014}\\
 & TRF2 & non-Fickian & multiple & sliding & $\bigcirc$ & 0.5~s $\sim$ 100~s & \cite{lin2014}\\
 & SA1 & non-Fickian & multiple &  &  & 0.5~s $\sim$ 40~s & \cite{lin2016}\\
\hline
\multicolumn{1}{c}{polymerase}  & T7 RNA polymerase & non-Fickian &  & sliding &  & 0.2~s $\sim$ 20~s & \cite{kim2007, barbi2004, barbi2004model}\\
\hline
\multicolumn{1}{c}{PARP}  & PARP1 & mixed & multiple &  &  & 0.1~s $\sim$ 50~s & \cite{liu2017b}\\
\hline
\multicolumn{1}{c}{TAL effector}  & TAL effector & Fickian & multiple & mixed & $\bigtimes$ & 20~ms $\sim$ 5~s & \cite{cuculis2016, cuculis2015}\\
\hline
\end{tabular*}
}
\end{table*}


\section{Diffusion of DNA-binding proteins: Fickian diffusion ($\alpha=1$)}
\label{sec:fickian}


\subsection{Tumor suppressor p53}
\label{sec:p53}
Tumor suppressor p53 is one of the essential transcription factors responsible for the repression of cancer formation. This protein is activated in response to stress signals such as DNA-damage signals, oxidative stress, and osmotic shock, inducing cell cycle arrest or apoptosis by transactivating its target gene\cite{kruse2009} to repair the damaged DNA or discard the damaged cell~\cite{brown2009, beckerman2010, bieging2014}. To transactivate its target genes, p53 should recognize them through a suitable diffusion process.

The homo tetrameric protein p53 has four identical subunits, each composed of an N-terminal domain (NTD), core, tetramerization (TET), C-terminal domain (CTD), and the linker connecting the core and the TET domain~\cite{joerger2008}. Among these, p53 uses two of them as the DNA-binding domains: the core domain for a specific binding to its target sites and the CTD for the nonspecific binding~\cite{kitayner2010, weinberg2004}. This multi-domain interaction, as described later, leads to heterogeneous multi-mode dynamics~\cite{tafvizi2011p53, murata2015, murata2017, subekti2017, kamagata2017}. Interestingly, despite its diverse DNA-protein interactions, p53 always shows Fickian diffusion dynamics with the anomaly exponent $\alpha=1$ at the time scale of $0.05$ to $0.5~\mathrm{s}$~\cite{tafvizi2011p53, murata2015, murata2017}.

To understand the multi-mode dynamics of p53, we first have to understand how each domain interacts with the DNA strands. p53 truncation mutants were conventionally used to study the dynamics of each domain. The mean diffusion constants of a full-length p53 ($D\sim0.16~\mu\mathrm{m}^{2}/\mathrm{s}$) and its truncation mutant containing the C-terminal domain (hereafter, TC) ($D\sim0.7~\mu\mathrm{m}^{2}/\mathrm{s}$) are independent of the ionic strength of $\mathrm{K}^{+}$ in the presence of $2~\mathrm{mM}~\mathrm{Mg}^{2+}$. On the other hand, the mean diffusion constant of the truncation mutant containing the core domain (hereafter, NCT) monotonically increases from $0.02~\mu\mathrm{m}^{2}/\mathrm{s}$ to $0.06~\mu\mathrm{m}^{2}/\mathrm{s}$ as the ionic strength increases from $25$ to $175~\mathrm{mM}~\mathrm{K}^{+}$ in the presence of $2~\mathrm{mM}~\mathrm{Mg}^{2+}$~\cite{tafvizi2011p53}. The small diffusion constant and the kymograph of the core domain in \cite{tafvizi2011} imply that the core domain is effectively immobile and bound to a nonspecific site of the DNA. To summarize, the C-terminal domain slides along the DNA with a smooth energy landscape with continuous contact, while the core domain repeatedly attaches to and detaches from the DNA, feeling a rugged energy landscape.

Due to the multi-domain interaction, the full-length p53 containing both domains for its search exhibits a non-Gaussian displacement distribution fitted with multiple Gaussian curves, while its C-terminal domain mutant exhibits a single-Gaussian displacement distribution~\cite{murata2015, murata2017, subekti2017}. It was suggested that there are three distinct diffusion strategies in this heterogeneous diffusing dynamics depending on the number of core domains 
attached to the DNA: (1) a slow mode where all parts of the C-terminal and core domains are in contact with the DNA strands, (2) a fast-I mode where only a small number of the core domains are in contact with the DNA, and (3) a fast-II mode where only the C-terminal domain is in contact with the DNA~\cite{murata2017}. The protein in the slow mode feels a highly rugged energy landscape due to the core domains, and translocates slowly. The p53 protein in the fast modes moves more freely since fewer core domains are interacting with the DNA, feeling a less rugged energy landscape. For all cases, however, the diffusion is always Fickian and does not exhibit subdiffusion regardless of the ruggedness of the energy landscape.

A coarse grained simulation study supports this multi-mode hypothesis~\cite{terakawa2012}. In this study, the core domains and C-terminal domains repeatedly dissociate from and reassociate to the DNA, changing the number of DNA-contacting domains stochastically. Most of the core domain and the C-terminal domain is in contact with the DNA at a low ionic concentration, while the rates of contacting domains decrease at higher ionic concentrations. It has been noted that the detached domains do not go far away from the DNA since one of the core domains or the C-terminal domains are still in contact with the DNA. The dissociated core domains escape $60~\mathrm{nm}$ away from the DNA due to the long linker connecting the core domain and the tetramerization domain, and the C-terminal domains maintain a close distance to the DNA even when it dissociates from the DNA. This stochastic change of the contacting domains can lead to transitions between the diverse search modes.

\subsection{Architectural DNA-binding proteins Nhp6A, HU, and Fis}

Architectural DNA-binding proteins assist the biological activities of other DNA-binding proteins by inducing a conformational change of DNA. One of the architectural proteins, HMGI(Y), is known to accelerate the binding of transcription factors by antagonizing intrinsic distortions of AT-rich DNA, and HMG14 and 17 are known to induce a moderate destabilization of the chromatin structure to facilitate transcriptional activation~\cite{bustin1996, bustin1999}. Among these architectural proteins, here, we focus on the diffusion dynamics of Nhp6A, HU, and Fis.
In \cite{kamagata2018}, the MSDs of these proteins were measured using a single-molecule fluorescence imaging. They are linear in the time window from 0.1~s to 0.35~s. The diffusion constants were estimated to be $0.33~\mu\mathrm{m}^{2}/\mathrm{s}$ for Nhp6A, $0.45~\mu\mathrm{m}^{2}/\mathrm{s}$ for HU, and $0.15~\mu\mathrm{m}^{2}/\mathrm{s}$ for Fis. It was also found that the diffusion constants do not vary significantly upon a change of salt concentration from $50$ to $200~\mathrm{mM}$ potassium glutamate (KGlu).

Nhp6A and HU also show a Gaussian displacement distribution $P(x, t)$ and their shapes do not change significantly under various ionic strength conditions. Fis, on the other hand, shows a non-Gaussian displacement distribution $P(x, t)$ due to its two-state dynamics. $90\%$ of the molecules were immobile and only $10\%$ were moving along the DNA arrays, and transitions between the two states was observed within the evaluated time interval ($\sim1~\mathrm{s}$). The average diffusion constants for the mobile (or fast) and immobile (or slow) molecules are $0.19\pm0.02$ and $0.007\pm0.006~\mu\mathrm{m}^{2}/s$, respectively, in the $150~\mathrm{mM}~\mathrm{KGlu}$ buffer condition.

It was noted that the diffusion constants of the architectural DNA-binding proteins are usually smaller than those of other DNA-binding proteins of comparable size (e.g., $D\sim2.2~\mu\mathrm{m}^{2}/\mathrm{s}$ for AVP-pVIc). A speculation is that the architectural DNA-binding proteins perform a rotation-coupled translocation along the DNA helical structure with a high free-energy barrier.
Eq.~\eqref{eqn:helical} indicates that the architectural DNA-binding proteins feel higher free-energy barrier ($\sim 1.7 k_{B}T$) than the other proteins ($\sim 1 k_{B}T$). This high free-energy barrier might be due to the large conformational changes of the DNA structure that these proteins induce. Binding of the architectural DNA-binding proteins accompanies variation in groove widths, bending of the DNA helical axis, and alterations in the base pair twist. Indeed, the architectural DNA-binding proteins induce greater DNA bending angles than the other non-architectural proteins with smaller free-energy barriers~\cite{kamagata2018}. A coarse-grained MD simulation of architectural DNA-binding protein HU supports this idea~\cite{tan2016}. This study showed that the diffusion of HU is highly coupled with DNA bending. Both of the cases were observed in this study: HU binding induced a sharp DNA bending and the bent DNA structure enhances the HU binding. The HU translocation was slower at the bent position of DNA and it recovers a faster diffusion at the less-bent position. It was found that HU binds to the concave side of bent DNA on which the negative charge density is higher than on the convex side. The stronger electrostatic interaction between HU and the bent side of DNA might induce the higher free-energy barrier for 1D translocation. It was noted that the high free-energy barrier of the architectural DNA-binding proteins does not result in sub-diffusive dynamics for these proteins~\cite{kamagata2018}.

\subsection{Mismatch repair proteins}

The mismatch repair proteins MutS and MutL recognize mismatched sites and initiate a mismatch repair (MMR) process. MutS homologs seem to form two kinds of clamps: an ADP-bound searching clamp and an ATP-bound sliding clamp. Once a MutS homolog recognizes its target site, ADP is exchanged with the MutS to form an ATP-bound clamp. After forming an ATP-bound sliding clamp, MutS escapes from the mismatch and diffuses freely along the DNA~\cite{hanaoka2016, cho2012, gorman2007}. The diffusion dynamics of an MutS was investigated in \cite{cho2012} using the technique of FRET and a Cy3-labelled MutS. This study revealed that both sliding MutS and ATP-bound MutS exhibit Fickian diffusion when they are not bound to a mismatch site and the displacement distribution follows a Gaussian distribution.  
Although both of them have similar Brownian diffusion, their dependence on the salt concentration is different, which is a typical clue for distinguishing between hopping and sliding.

In the case of the ADP-bound MutS, the diffusion constant, $D\sim 0.032\pm0.001 ~\mu\mathrm{m}^{2}/{s}$, is independent of the salt concentration, which suggests that the ADP-bound MutS slides on the DNA with a continuous contact. Its dwell time during which an MutS is bound to the DNA decreases from $7$~s to $1$~s as the salt concentration increases from $25$ to $150$ $\mathrm{mM}~\mathrm{NaCl}$. On the other hand, the ATP-bound MutS has a dwell time $=683\pm22~s$ independent of the salt concentration, and its diffusion constant increases from $0.05$ to $0.17~\mu\mathrm{m}^{2}/\mathrm{s}$ as the salt concentration increases from $25$ to $300~\mathrm{mM}~\mathrm{NaCl}$. This implies that the ATP-bound moves along the DNA with a discontinuous contact and forms a more stable structure so that it has a longer dwell time.

Both ADP-bound and ATP-bound MutS proteins were found to rotate around the DNA helical axis by smPolarization-TIRF method~\cite{cho2012}. The ADP-bound searching MutS displays rotational-coupled sliding dynamics and the ATP-bound MutS freely rotates around the DNA helical axis: the rotation was significantly restricted when a MutS moves along a short DNA segment with an effective diffusion length of $\sim5~\mathrm{bp}$. This result supports the idea that the rotational motion of ADP-bound MutS is coupled with the sliding motion. On the other hand, the rotation of an ATP-bound MutS was independent of the length of the DNA segment, which suggests that the ATP-bound MutS performs hopping.

The diffusion dynamics for several homologs of MutS were also investigated experimentally. Msh2-Msh6, the \textit{S.cerevisiae} homolog of MutS, has dynamics similar to MutS. It slides along the DNA in continuous contact, and its diffusion constant is not affected by the ionic strength~\cite{brown2016}. Its other homolog Msh2-Msh3, on the other hand, exhibits quite different dynamical properties~\cite{brown2016}. Msh2-Msh3 shows Brownian dynamics with $\alpha \sim 1$ and a Gaussian distributed displacement distribution, which is the same as the other homologs (MutS and Msh2-Msh6). Unlike ADP-bound MutS and Msh2-Msh6, which slide along the DNA with a continuous contact~\cite{cho2012, gorman2007}, the ADP-bound Msh2-Msh3 seems to use the hopping to move along the DNA. A diffusion constant that depends on the salt concentration is evidence of hopping. As the ionic strength increases from $25$ to $150~\mathrm{mM}~\mathrm{NaCl}$, the effective diffusion constant of ADP-bound Msh2-Msh3 increased from $(3.1\pm2.7)\times 10^{-2}~\mu\mathrm{m}^{2}\mathrm{s}^{-1}$ to $(1.2\pm1.4)\times10^{-1}~ \mu\mathrm{m}^{2}\mathrm{s}^{-1}$.  Moreover, Msh2-Msh3 in $1~\mathrm{mM}$ ADP and $100~\mathrm{mM}~\mathrm{NaCl}$ buffer condition could transfer between two closely neighboring extended DNA substrates, which are separated by $1~\mu\mathrm{m}$. It can also occasionally bypass protein obstacles on the DNA, independently of the nucleotide attached to the Msh2-Msh3. These properties distinct from the other MutS homologs are revealed to originate from the Msh3-specific mispair-binding domain (MBD). By replacing the Msh6 MBD in Msh2-Msh6 with Msh3 MBD, Msh2-Msh6 with Msh3 MBD(Msh2-Msh6(3MBD)) showed Msh2-Msh3-like dynamics. Increasing the ionic strength from $50$ to $150~\mathrm{mM}~\mathrm{NaCl}$ with $1~\mathrm{mM}~\mathrm{ADP}$ increased the average diffusion constant of Msh2-Msh6 (3MBD) from $(5.8\pm5.2)\times10^{-2}~\mu\mathrm{m}^{2}s^{-1}$ to $(1.9\pm1.1)\times10^{-1}~\mu\mathrm{m}^{2}s^{-1}$. Furthermore, replacing the MBD enabled Msh2-Msh6(3MBD) to bypass nucleosomal encounters on the DNA and to transfer between neighboring DNA substrates, with almost the same frequencies as Msh2-Msh3~\cite{brown2016}.

Mlh1-Pms1 is a yeast and the human homolog of MutL in \textit{E. coli}, which processes the mismatch repair process after MutS homologs recognize the mismatch. It forms a ring-like structure with a central pore of $16.5\pm4.6~\mathrm{nm}$ on a diameter through the protein--protein interactions between the C-terminal domains and the N-terminal domains, the C-terminal and N-terminal domains being connected by linker arms. It is believed that this ring structure encloses the DNA when it moves along the DNA with discontinuous contact, which would lead the hopping dynamics and thermal-driven Fickian dynamics. The MSD of Mlh1-Pms1 has an almost linear relation with the lag time and has a diffusion constant that is dependent on the salt concentration, implying the hopping dynamics~\cite{gorman2010}. As the salt concentration increases from $25~\mathrm{mM}~\mathrm{NaCl}$ to $200~\mathrm{mM}~\mathrm{NaCl}$, the diffusion constant increases from $0.026\pm0.017~\mu\mathrm{m}^{2}/\mathrm{s}$ to $0.99\pm0.411~\mu\mathrm{m}^{2}/\mathrm{s}$. Bypassing obstacles is another expected feature for hopping of this ring-like protein. Mlh1-Pms1 can bypass another Mlh1-Pms1, nucleosomes on DNA, and sometimes, although with less frequency, bypass Qdot-tagged nucleosomes which have a larger diameter than the central pore of the Mlh1-Pms1. These results suggest that Mlh1-Pms1 sometimes opens its ring-like structure to bypass a large obstacle during 1D translocation.

\subsection{Human oxoguanine DNA glycosylase 1~(hOGG1)}
Human oxoguanine glycosylase 1~(hOGG1) is a DNA glycosylase enzyme that is involved in the excision of 8-oxoguanine, the lesion site resulting from reactive oxygen species. In \cite{blainey2006} it was found experimentally that hOGG1 performs Fickian diffusion along the DNA at time scales of $0.05~\textrm{s}$ to $1.6~\mathrm{s}$ under various pH and salt concentrations. Its diffusion constant is, however, significantly affected by the pH~\cite{blainey2006}. As the pH increases from 6.6 to 7.9, the diffusion constant grows by more than an order of magnitude. On the other hand, varying the salt concentrations does not change the effective diffusion constant, which points towards sliding dynamics. Indeed, changing the size of a dye attached to a hOGG1 protein led to finding that the diffusion constant follows the rotation-coupled translation model given by Eq.~(\ref{eqn:helical})~\cite{blainey2006, blainey2009}.

In recent work, hOGG1 was found to have two states: a loosely bound state and a tightly bound state~\cite{vestergaard2018}, who measured both the diffusion constant and the residence time of hOGG1 on the DNA substrates. If hOGG1 had just one state, its diffusion constant should be constant, regardless of the residence time. However, the experimental result show that the diffusion constant is dependent on the residence time: particles with a short residence time have higher diffusion constants, which implies that there are multiple kinetic states for the diffusion. They proposed a simple two-state model to explain the relation between the diffusion constant and the residence time. In this model, a particle switches between the two states. Each state has its own diffusion constant, dissociation rate, and transition rate. The prediction of the experimental results by the two-state model was much better than the one-state model. In this two-state model, the estimated diffusion constant for the fast mode is more than ten times faster than those for the slow mode, and the proteins spend roughly equal times in each state~\cite{vestergaard2018}.

\subsection{EcoRV}
EcoRV is a type II restriction endonuclease, which recognizes its target site, a palindromic 6-base DNA sequence, to produce a cleavage in it. In a TIRF-based experiment~\cite{dikic2012}, the MSD curves of EcoRV with different sizes of fluorescence dyes (scRM6, Cy3B, savCy3, QEDO6, QD605 PEG11, QD605 PEG2, and QD655) were obtained. Each MSD curve exhibits a linear dependence with the time lag, but they have different diffusion constants. The friction coefficients were estimated from the diffusion constants, which showed a size dependence of $R^{3}$~($R$ is the radius of each fluorescence dye). This result suggests that an EcoRV undergoes a rotation-coupled sliding along the DNA, as we have seen in Eq.~(\ref{eqn:helical})~\cite{dikic2012}.

In another TIRF experiment with non-cognate DNA substrates, a Cy3B-labelled EcoRV frequently made large jumps over $200~\mathrm{nm}$ in $20~\mathrm{ms}$ between Fickian translocations with $\alpha=1$ along the DNA~\cite{bonnet2008, biebricher2009}. These long jumps are probabilistically hard to observe for the obtained diffusion constant $D\sim10^{-2}~\mu\mathrm{m}^{2}/\mathrm{s}$ in a 1D Brownian motion.

To see whether the large jumps are made by 3D diffusion or 1D sliding, the effect of the hydrodynamic flow and the ionic strength on the jump dynamics were studied. When a hydrodynamic flow was applied to EcoRV, the total number of large jumps decreased and the overall jump lengths also decreased. Similar results were also observed when the ionic strength increased. As the ionic strength increased from $10$ to $60~\mathrm{mM}~\mathrm{NaCl}$, the number of long jumps decreased. It is speculated that the hydrodynamic flow and the screening effect by a high salt concentration make it hard for the protein to re-associate after a long jump. Furthermore, they showed that a simple numerical simulation where a protein can dissociate from and re-associate to the DNA with a fixed probability reproduces the same cumulative length distribution of the large jumps~\cite{bonnet2008}. These results support the idea that the large jumps are performed by 3D diffusion.

The large jumps mentioned above were defined as those with a step length longer than $200~\mathrm{nm}$ per unit time. It turned out that there are also short jumps (or hopping) that cannot be distinguished from 1D sliding with continuous contact with the DNA. If the diffusion constant depends on the salt concentration, that supports the existence of a hopping-based translocation. As the ionic strength increases from $10$ to $60~\mathrm{mM}~\mathrm{NaCl}$, the average diffusion constant also increases from $1.1\pm0.1\times10^{-2}$ to $2.5\pm0.1\times10^{-2}~\mu\mathrm{m}^{2}/\mathrm{s}$.

In another experiment~\cite{biebricher2009}, it was found that although EcoRV partly uses hopping dynamics, the translocation occurs mainly along the DNA contour. They tracked the QD-labelled EcoRVs moving along a DNA molecule held in an extended conformation. Using optical tweezers, the DNA extension was controlled from $65\%$ to $105\%$ of the DNA contour length. As the extension increased from $65\%$ to $100\%$ of the DNA contour length, the diffusion constant of the EcoRV also increased. After correcting the MSD curves considering the extensions, the diffusion constants were found to have almost the same values, $3.17\pm0.23\times10^{-3}~\mu\mathrm{m}^{2}/\mathrm{s}$, $3.01\pm0.2\times10^{-3}~\mu\mathrm{m}^{2}/\mathrm{s}$, and $3.15\pm0.15\times10^{-3}~\mu\mathrm{m}^{2}/\mathrm{s}$ for the extensions at $65\%$, $80\%$, and $100\%$, respectively. This result implies that the translocation occurs mainly along the DNA contour, not by inter-segmental transfer. When the extension becomes higher than $100\%$, the corrected diffusion constant becomes smaller than those for extensions less than $100\%$, suggesting that the DNA deformation changes the energy landscape.

\subsection{LacI}
Lac repressor~(LacI) inhibits the gene expression for lac proteins that are involved in the metabolism of lactose, by binding to the lac operon. LacI is the first protein whose association rate was found to be 100 times faster than those of the 3D diffusion limit~\cite{riggs1970}. In experiments based on TIRF~\cite{wang2006}, a LacI moving along the noncognate sites of DNA shows a Fickian dynamics with a broad range of $D$. The diffusion constant of each protein ranges from $2.3\times10^{2}~\mathrm{nm}^{2}/\mathrm{s}$ to $1.3\times10^{5}~\mathrm{nm}^{2}/\mathrm{s}$. The displacement distributions $P(x, t)$ for the first 15 points appear to be Gaussian despite the broad range of $D$, which could be due to the lack of the sample steps in the experiment~\cite{lampo2017}.

The diffusion length $l_{d}$ before dissociation was also obtained experimentally to estimate the acceleration of the target binding rate of the Facilitated Diffusion. Using the experimentally obtained quantities (mean diffusion length $\langle l_{d} \rangle\approx 500~\mathrm{nm}$, 1D diffusion constant $\langle D\rangle\approx2.1\times10^{-10}~\mathrm{cm}^{2}$), and the given quantities (3D diffusion constant $D_{3d}\approx 4\times10^{-7}~\mathrm{cm}^{2}/\mathrm{s}$, the total length of the DNA molecule $L=15.5~\mu\mathrm{m}$, target concentration $c\approx1~\mathrm{lacO}/1670~\mu\mathrm{m}^{3}$, and effective DNA target length $l_{\mathrm{seq}}\approx3~\mathrm{bp}$), the acceleration factor for the Facilitated Diffusion can be calculated by
\begin{eqnarray}
k_{a}&=&D_{3d}l_{\mathrm{seq}} \left(\frac{l_{\mathrm{seq}}}{l_{d}}+\frac{D_{3d}}{D}l_{\mathrm{seq}}Ll_{d}c\right)^{-1} \\
&\approx& 100 D_{3d}l_{\mathrm{seq}},
\end{eqnarray}
which is about 100 times faster than the 3D diffusion limit. The experimental result of 1D dynamics thus supports the faster-than-diffusion association rate observed in the Riggs' experiment.

In a recent study using smFRET and single-molecule confocal laser tracking combined with fluorescence correlation spectroscopy (SMCT-FCS)~\cite{marklund2020}, LacI was found to show a rotation-coupled sliding and occasional hopping. The autocorrelation function of the fluctuation in the fluorescence signal was explained well by rotation-coupled translocation. The protein completes one revolution after traversing about 40~bp, which exceeds the 10.5~bp helical pitch of DNA. This implies that LacI frequently hops out of the DNA groove while it moves along the DNA. Frequent bypassing events over its operator sequences ($>90\%$) also support there being a hopping dynamics~\cite{hammar2012, marklund2020}. In spite of the frequent hopping, its translocation seems to be obstructed to some extent by the protein roadblocks (i.e., the transcription factor TetR)~\cite{hammar2012}. The overall results support the idea that LacI uses both rotation-coupled sliding and hopping during 1D translocation.

\subsection{Rad51}
Rad51 is involved in the homologous recombination repair process. A single-molecule study using TIRF~\cite{graneli2006} showed that a Rad51 performs a one-dimensional Fickian dynamics at $0.1~\mathrm{s}\sim 10~\mathrm{s}$ and tightly binds to the free ends of the DNA. This finding suggests that Rad51 non-specifically binds to the DNA strands and undergoes a conformational change at the ends. In the absence of a hydrodynamic force, on the other hand, Rad51 shows unbiased thermal-driven dynamics.

There is evidence to suggest that the mobile Rad51 is in a ring-like conformation and encircles the DNA during the diffusion: Rad51 stably binds on the DNA for up to several hours, and no bypassing event between two Rad51 was observed. Thus it is suggested that Rad51 makes close contact with the DNA in a ring-shaped conformation enclosing the DNA substrate.

\section{Diffusion of DNA-binding proteins: anomalous diffusion ($\alpha\neq1$)}
\label{sec:anomalousproteins}
In this section, we review the diffusion dynamics of DNA-binding proteins that were found to have non-Fickian characteristics. In this  case, the MSD of these proteins often scales as $\langle x^2(t) \rangle \propto t^\alpha$ with $\alpha\neq 1$. Since the diffusion is not Fickian, the diffusion constant $D$ for a non-Fickian process is usually not defined; see Section~III for further information on the generalized diffusion constant $D_\alpha$ for anomalous diffusion processes. Nevertheless, in many experimental studies, one measures an \emph{effective} diffusion constant $D_{\mathrm{eff}}$ in the sense of linear regression, i.e., 
\begin{equation}
\langle x^2(t)\rangle =2D_\alpha t^\alpha=2 D_{\mathrm{eff}}(t) t.
\end{equation}
Note that the estimated $D_{\mathrm{eff}}\sim t^{\alpha-1}$ ($\alpha<1$) changes with the measurement time $t$ (i.e., the number of data points used). Hence, it is not physically meaningful to interpret the value of $D_{\mathrm{eff}}$ itself. Instead, one can get information about the anomaly ($\alpha$) of the diffusion process by comparing $D_{\mathrm{eff}}$ for given $t$. In the review that follows, we distinguish the effective diffusion constant $D_{\mathrm{eff}}$ from the genuine diffusion constant $D$ defined in Fickian diffusion, and consistently use the notation $D_{\mathrm{eff}}$ for the diffusion constants of non-Fickian protein motion reported in the literature.

\subsection{Rad4-Rad23, XPC-Rad23B}
Xeroderma pigmentosum group C (XPC) and its yeast ortholog Rad4-Rad23 are known to recognize a wide variety of damage in DNA substrates and initiate the DNA-repair process~\cite{min2007, chen2015, cheon2019, sugasawabook2019}. Rad4-Rad23 has three distinct diffusion patterns, each of which has its own distinct diffusion constant and anomaly exponent~\cite{kong2016}. These three dynamic states are called non-motile, random, and constrained motion. Constrained motion is, by definition in this review, classified as a subdiffusive 1D motion along the DNA ($\alpha<1$). Random motion is the case that the XPC molecules have a Fickian diffusion with $\alpha\approx 1$. Non-motile motion is the case where the displacement of the XPC molecule on a DNA is indistinguishable from the drift motion of the DNA itself~\cite{kong2016, cheon2019}. It was found that the constrained motion is not affected by the salt condition in the sense that the anomaly measures, $\alpha$ and $D_\alpha$, are independent of the salt concentration. For the XPCs performing a Fickian diffusion, on the other hand, the diffusion constant increases with increasing salt concentration.

XPC-Rad23B, the human ortholog of Rad4-Rad23, exhibits analogous diffusion dynamics with Rad4-Rad23 in that the protein dynamics is classified into three types, depending on the spatial range of displacement for a given measurement time~\cite{cheon2019}. For both non-cognate and CPD-containing $\lambda$-DNAs, the XPC-Rad23B molecule has immobile, diffusive, and constrained motion. For the given measurement time of 5 min, the constrained proteins move within a couple of thousand base pairs. The immobile proteins seem to be strictly bound to the DNA. The protein in the diffusive mode is shown to move freely along the DNA with the highest diffusion constant. Importantly, even the diffusive XPCs were found to follow a subdiffusion with the anomaly exponent $\alpha < 1$ in the time window of $[0.05~\mathrm{s},~ 15~\mathrm{s}]$~\cite{cheon2019}. The exact anomaly exponents are not considered here, but the MSD curves evidently show subdiffusive features both for the diffusive proteins and the constrained proteins. Interestingly,  the population of the constrained proteins is highly correlated with the  AT-rich regions. From this observation, the authors speculated that the AT-rich regions allow the local opening of the DNA for the XPCs to be easily trapped at these sites~\cite{cheon2019}.

The three dynamical states behave differently to changes in the salt concentration. Increasing the salt concentration from $40$ to $150~\mathrm{mM}~\mathrm{NaCl}$ caused a nearly 10-fold increase in $D$, from $0.034\pm0.045~\mu\mathrm{m}^{2}/\mathrm{s}$ to $0.39\pm0.23~\mu\mathrm{m}^{2}/\mathrm{s}$ for the diffusive motion. On the other hand, constrained proteins and immobile proteins did not show a significant change in their dynamics with varying salt concentration~\cite{cheon2019}. From this result, it was suggested that the XPC in the diffusion mode moves along the DNA via a hopping mechanism. Another piece of experimental evidence supporting there being hopping dynamics is that the protein can bypass the protein roadblocks (EcoRI) on the DNA~\cite{cheon2019}. The authors used an EcoRI mutant which binds to the DNA tightly as a protein roadblock. When an XPC encountered the protein roadblock, it bypassed the roadblock with a 30\% to 40\% probability. Based on this observation, it was proposed that the XPC diffusion is based on hopping.

\subsection{Telomeric sequence binding proteins TRF1, TRF2, and SA1}
TRF1 and TRF2 are the subunits of the shelterin protein complex, which is known to protect telomeres from the DNA repair mechanism by preventing non-homologous end joining, end-to-end fusion, and apoptosis~\cite{sfeir2012}. TRF1 and TRF2 directly bind onto the double-stranded telomeric sequences, thus playing a critical role in the target search mechanism of the shelterin complex. In \cite{lin2014}, the kymographs of TRF1 and TRF2 subunits moving along the DNA were obtained using \emph{in vitro} DNA tightrope assay. Both TRF1 and TRF2 follow a subdiffusive dynamics, but they respond differently to the ionic strength on the nontelomeric DNA substrate where no binding sites exist. Upon an increase of ionic strength from $75$ to $225~\mathrm{mM}D$($50~\mathrm{mM}$ HEPES and NaCl ($25$, $50$, $75$, and $100~\mathrm{mM}$)), the $D_{\mathrm{eff}}$ of TRF1 decreases from $(7.5\pm1.2)\times10^{-2}~\mu\mathrm{m}^{2}/\mathrm{s}$ to $(3.8\pm1.2)\times10^{-2}~\mu\mathrm{m}^{2}/\mathrm{s}$ while the anomaly exponent increases from $\alpha\approx 0.65$ to $\alpha\approx0.89$. On the other hand, the diffusion dynamics of the TRF2 protein turns out to be insensitive to the salt concentration. The value of $D_{\mathrm{eff}}$ is consistently $\sim9\times10^{-2}~\mu\mathrm{m}^{2}/\mathrm{s}$ and $\alpha$ is about $0.9$.

It was also investigated whether these proteins can bypass a protein roadblock bound to the DNA substrates. It turns out that neither of them bypasses the obstacle, so hopping is not the mechanism of the 1D diffusion. The experiment further showed that the $D_{\mathrm{eff}}$ of these proteins are consistent with those predicted for an object rotating around the DNA, which is given by Eq.~\eqref{eqn:helical}~\cite{blainey2009}. Collecting the observed properties for their diffusion dynamics, it can be inferred that TRF2 is a canonical slider while TRF1 seems to slide with a putative conformational change depending on the ionic concentration.

An additional important feature for their diffusion dynamics is that both proteins have a sequence-sensitive diffusive motion while moving along the telomeric sequences~\cite{lin2014}. The diffusion constant inside the telomeric sequences was found to be 10 to 30 times slower than for the non-telomeric sequences. These proteins also show confined motion within these repeated target sequences. Within the target regions, both proteins seem to feel a lower energy landscape compared to the non-telomeric sites.

SA1 is a subunit of the cohesin complex, required for telomere cohesion~\cite{bisht2013}. In a tightrope assay~\cite{lin2016}, SA1 was found to have sequence-dependent two-state dynamics: fast and slow diffusion. Within the DNA segments containing telomeric sequences, SA1 continuously alternates between the fast and slow diffusions whereas within the genomic or centromeric sequences only a few populations show dynamics.

The slow diffusion was revealed to be mediated through the N-terminal domain containing an AT-hook motif whose C-shaped structure can be inserted into the DNA minor groove, making hydrogen bonds or electrostatic interaction with the DNA phosphate groups. To see whether this N-terminal leads to the slow dynamics, they observed the dynamics of the N-terminal fragment and found a similarly fast and slow alternating dynamics.

The pausing events during the free diffusion seem to lead to a more subdiffusive dynamics with an anomalous exponent of $\alpha\sim0.7$ at the telomeric sequences, whereas it is less subdiffusive at genomic ($\alpha\sim0.89$) and centromeric ($\alpha\sim0.82$) sequences.

\subsection{Glycosylase family Fpg, Nei, and Nth}
DNA glycosylases involved in base excision repair process find a damaged site and remove the damaged bases to initiate the base excision repair process. It was found that DNA glycosylases of E. Coli such as Fpg, Nei, and Nth have very similar diffusion dynamics~\cite{dunn2011, nelson2014}. The effective diffusion constants $D_{\mathrm{eff}}$ for these glycosylases range from $0.001~\mu \mathrm{m}^{2}/\mathrm{sec}$ to $1~\mu \mathrm{m}^{2}/\mathrm{s}$, and the $\alpha$ from $0.1$ to $1.4$, respectively, with a positive correlation. The distributions of $D_{\mathrm{eff}}$ and $\alpha$ were found to be insensitive to the salt concentration, while the lifetime of the binding of the glycosylases decreases as the concentration of potassium glutamate (KGlu) increases from 0 to 250~mM.

The subdiffusive and slow dynamics seems to originate from the amino acid residues of the glycosylases. It is known that the glycosylases use their wedge residues to interrogate for the damaged site. They insert the wedge residues into the intrahelical structure to test the strength and flexibility for given base pair~\cite{banerjee2006, qi2010}. In \cite{nelson2014, dunn2011}, the diffusion dynamics of wedge mutants, whose phenylalanine wedge residues of Fpg, Nei, and Nth are replaced by alanine, was studied~\cite{dunn2011, nelson2014}. These wedge mutants have larger $D_{\mathrm{eff}}$ compared to their wild-type ones, both on the undamaged DNA substrates and the damaged DNA substrates containing the oxoG lesion sites. In the case of Fpg, not only did they become faster, but their subdiffusive populations also disappeared. This result suggests that the interrogation process by the wedge residue is responsible for the subdiffusive and slow dynamics of Fpg, Nei, and Nth.

The wedge residue was not the only origin of the slow dynamics of the glycosylases. The number of oxidatively damaged bases also affects their diffusion constants. As the number of damaged sites increases, the $D_{\mathrm{eff}}$ of the glycosylases monotonically decreases even when the phenylalanine wedge residues are replaced by alanine. In \cite{nelson2014}, it was suggested that a glycosylase frequently pauses to interrogate the DNA by inserting its wedge residue. If the given base is undamaged, it soon resumes scanning. On the other hand, if the given base is damaged, it readily binds to the damaged site, which complicates the lesion recognition. Even though the phenylalanine residue is replaced by alanine, glycosylase is still able to interrogate DNA bases with a lower efficiency.

Another interesting property of the glycosylases was the broad range of $D_{\mathrm{eff}}$ ($\sim0.001~\mu\mathrm{m}^{2}/\mathrm{s}$ to $\sim1~\mu\mathrm{m}^{2}/\mathrm{s}$), which suggested
that there are several diffusion states. They proposed three possible types of diffusive motion based on the order of $D_{\mathrm{eff}}$. Particle motion with $D_{\mathrm{eff}}\sim 0.001~\mu m^{2}/s$ is indistinguishable from the DNA fluctuation and thus considered to be a state bound to the DNA substrates.
For proteins with $D_{\mathrm{eff}}\sim0.01~\mu m^{2}/s$, they were assumed to undergo a rotation-coupled diffusion along the DNA helical structure. In this case, the relatively smaller diffusion constant compared to the expected value from Eq.~\eqref{eqn:helical} ($D\sim0.05~\mu\mathrm{m}^{2}/\mathrm{s}$) might be due to frequent pausing to insert the wedge residue into the bases to interrogate damage. Glycosylases with $D_{\mathrm{eff}}\sim 0.1~\mu \mathrm{m}^{2}/\mathrm{s}$ are faster than those undergoing rotation-coupled sliding. It was suggested that in this case the rotation-coupled and the rotation-decoupled translocation are combined.

\subsection{T7 RNA polymerase}
T7 is an RNA polymerase that recognizes T7 promoter and transcribes the DNA downstream of its promoter from $5'$ to $3'$ direction, powered by nucleotide triphosphate (NTP). It is known that T7 RNA polymerase forms hydrogen bonds with a series of DNA sequences~\cite{schick1995, li1996, cheetham1999, imburgio2000}. Based on this idea, Barbi \textit{et al}. introduced a sequence-dependent diffusion model, which predicted anomalous subdiffusion in the short-time regime~\cite{barbi2004, barbi2004model}. This model will be explained in Section~VI.

After this theoretical model study, the dynamics of T7 RNA polymerase was directly observed using TIRF microscopy. T7 proteins moving along a $16.4$-$\mu\mathrm{m}$-long $\lambda$-DNA without any T7 promoter sequence show sub-diffusive behavior in $10~\mathrm{s}$ with a non-Gaussian displacement distribution~\cite{kim2007}. The estimated $D_{\mathrm{eff}}$ ranged from $6.1\times10^{-11}~\mathrm{cm}^{2}/\mathrm{s}$ to $4.3\times10^{-9}~\mathrm{cm}^{2}/\mathrm{s}$. It was not affected by the salt concentration, suggesting that T7 maintains close contact with the DNA during diffusion. The broad range of $D_{\mathrm{eff}}$ is not due to statistical errors from lack of samples, since the simulated Brownian trajectories of the same length do not show such a large deviation. This broad range of diffusion constants might lead to the non-Gaussian displacement distribution that was shown in the experiment~\cite{kim2007, chechkin2017}.


\section{Theoretical models for protein diffusion}
\label{sec:theoreticalmodels}
\subsection{Two-state dynamics}
\label{sec:twostate}
A protein is required to find its target site sufficiently quickly, and at the same time, to stably bind to the target for its own biological function. However, the requirements of fast searching and stable binding are mutually exclusive. Fast diffusion is possible if the non-specific binding energies  between the protein and the DNA sequence are weak along the DNA. In terms of the (free) energy landscape that the DNA-binding protein feels when bound, the requirement is understood such that the energy landscape is sufficiently smooth where the energy deviation along the DNA sequence is about $\epsilon \sim 1\hbox{--}2~k_{B}T$. In contrast, $\epsilon$ should exceed $\sim 5~k_{B}T$ for the stable binding of the protein to the target site. This contradiction is referred to as the \textit{speed--stability paradox}~\cite{slutsky2004, mirny2009}.
This paradox can be resolved if a DNA-binding protein has multiple dynamic states. As a minimal model, the idea of two-state models has been proposed~\cite{slutsky2004, mirny2009}. In the two-state model, a protein generally has two distinct dynamic states (fast vs. slow) and switches between the two states during its target search process. On the one hand, the fast state describes a search mode where a protein freely diffuses along the DNA under a smooth energy landscape with a small value of $\epsilon/[k_BT](\sim1)$. On the other hand, the slow state mimics a recognition mode where a protein sensitively feels the rugged energy landscape (i.e., the local DNA sequence) with a high $\epsilon/[k_BT](\gtrsim 5)$, at which the protein dynamics may sensitively depend on the sequence. This is a short remark on the biological background for the introduction of the two-state models. Below we introduce a two-state diffusion model.  

Two-state diffusion dynamics can be understood in the framework of heterogeneous diffusion processes introduced in Section~III.~3. Let us assume that a DNA-binding protein has two distinct diffusion states having the diffusion constants $D_\pm$. In \cite{miyaguchi2016}, a two-state diffusion process was investigated in the framework of the Langevin equation \eqref{eqn:DDlangevin}, where the diffusivity $D(t)$ randomly fluctuates between two values, i.e., $D(t)=D_{+}$ (fast) and $D(t)=D_{-}$ (slow). The transition dynamics between $D_+$ and $D_-$ is described by a renewal process. The sojourn times of $D_\pm$ are governed by two PDFs $\rho_\pm(\tau)$, which  are given by a power-law distribution:\begin{equation}
\rho_{\pm}(\tau)\underrel{\tau \to \infty}{\simeq}\frac{a_{\pm}}{\abs{\Gamma(-\alpha_{\pm})}\tau^{1+\beta_{\pm}}}~\hbox{with}~\beta_\pm>0.
\end{equation}
Here, $\Gamma(x)$ is the Gamma function and $a_{\pm}$ is a scale factor. In the range of $\beta_\pm>1$, the mean sojourn time $\langle \tau\rangle_\pm=\int d\tau \rho_\pm(\tau)\tau $  for the state $D_\pm$ is finite; however, for a broad distributed PDF with $0<\beta_\pm<1$, the $\langle \tau\rangle_\pm$ diverges. 

The study showed that this two-state process has rich dynamic properties depending on the condition of $\beta_\pm$ and the initial sojourn time PDF $\rho_{\pm}^{0}(\tau)$ for the first transition event. Here, we are restricted to the case of the equilibrium ensemble (where $\beta_\pm>1$), which seems to be an appropriate condition for modeling the protein diffusion on the DNA. For the equilibrium ensemble, it is assumed that the renewal process for $D(t)$ starts $t=-\infty$ and a particle enters in the random environment (i.e., DNA in our problem) at $t=0$. Then, the probability for the particle starting from the state $D_\pm$ at $t=0$ is given by $p_{\pm}^\mathrm{eq}=\langle \tau\rangle_\pm/(\langle \tau\rangle_++\langle \tau\rangle_-)$~\cite{miyaguchi2019}, and the first sojourn time occurs with the PDF $\rho_{\pm}^{0}(\tau)=\rho_{\pm}^\mathrm{eq}(\tau)$ where  $\hat{\rho}_{\pm}^\mathrm{eq}(s)=(1-\hat{\rho}_{\pm}(s))/[\langle \tau\rangle_\pm s]$ in the Laplace domain (see \cite{miyaguchi2016} for the derivation). For the equilibrium ensemble, it was found that this process exhibits a Fickian dynamics with the MSD
\begin{equation}
\langle x^2(t) \rangle=2D_\mathrm{eq}t
\end{equation}
where the diffusion constant is given by
\begin{equation}
D_\mathrm{eq}\equiv \langle D(t) \rangle=D_+p_{+}^\mathrm{eq}+D_-p_{-}^\mathrm{eq}.
\end{equation}
Although the equilibrium process behaves like an ordinary Fickian motion in the MSD, the propagator $P(x,t)$ indicates that  the process is actually non-Gaussian. The propagator, in the Fourier-Laplace space, is
\begin{eqnarray}
\tilde{P}^\mathrm{eq}(k,s)&=&\frac{p_{\pm}^\mathrm{eq}}{s+D_\pm k^2} +\frac{[1-\tilde{\rho}_+(s_+)[1-\tilde{\rho}_-(s_-)]}{(s+D_\pm k^2)[1-\tilde{\rho}_+(s_+)\tilde{\rho}_-(s_-)]}\nonumber \\
&\times& \left(\frac{1}{s+D_\mp k^2}-\frac{1}{s+D_\pm k^2}\right)
\end{eqnarray}
where $\tilde{f}(k,s)=\int dx e^{-ikx}\int_0^\infty dt e^{-st}f(x,t)$. When  $D_+= D_-$ (i.e., $D(t)=D_+$ for all $t$), the second term on the R.H.S. is removed, $P^\mathrm{eq}(x,t)$ becomes Gaussian (as it should be). When $D$ has two states ($D_+\neq D_-$), however, the second term gives a non-Gaussian component to $P$. 
For any non-equilibrium ensemble, it turns out that the diffusion properties of the two-state process are quite complicated. The MSD and $P(x,t)$ have dramatically different behaviors, depending on the way of averaging as well as on the value of $\beta_\pm$. For further information about this case, refer to the original papers~\cite{miyaguchi2016,miyaguchi2019}.

\subsection{Diffusion on a sequence-dependent energy landscape}
In the previous section, the protein diffusion was modeled to simply have fast and slow states (or the search and recognition modes). While such two-state models provide an overview of the heterogeneous diffusion dynamics of a DNA-binding protein over a long time window, protein diffusion is usually more complicated than that. In particular, experimental studies have reported evidence that some proteins, e.g., T7 RNA polymerase~\cite{schick1995, li1996, cheetham1999, imburgio2000}, have sequence-dependent diffusion dynamics. 
A possible mechanism for sequence-dependent diffusion is the DNA--protein interactions for the target search. Namely, to bind onto the target sequence, the DNA-binding protein needs to interrogate whether a given sequence is the correct one or not, by forming hydrogen bonds perpetually. This interrogation and recognition process would make a protein feel a sequence-dependent rugged energy landscape. 

To study how this rugged energy landscape affects the diffusion dynamics of a protein, Barbi \textit{et al.} introduced a sequence-dependent protein diffusion model~\cite{barbi2004, barbi2004model}, where the major component of the DNA--protein interactions is the hydrogen bonding between the amino acids of the protein and the DNA base pairs. 
In this model, a protein is assumed to slide along the DNA from a site $n$ to its nearest neighbor sites $n\pm1$. Then, due to the sequence-dependent energy landscape, it feels a different effective energy barrier $\Delta U_{n\rightarrow n'}$ (where $n'=n\pm1$) over a diffusion jump. The diffusion rate to the next sites is given by~\cite{barbi2004model}
\begin{equation}\label{eq:jumprate}
r_{n\rightarrow n'}=\frac{1}{2\tau}\times \mathrm{exp}(-\Delta U_{n\rightarrow n'}/[k_{B}T]),
\end{equation}
where $\tau$ is the time that a protein needs to diffuse (or translocate) over the lattice unit between the sites (i.e., the base-pair distance). As seen in Eq.~\eqref{eq:jumprate}, in this model the diffusion dynamics is mostly governed by the energy barrier $\Delta U_{n\rightarrow n'}$.

There are four different models for $\Delta U_{n\rightarrow n'}$. (1) No-threshold model: The effective energy barrier is simply given by the energy difference between its original site and the nearest neighbor, or is zero if that is negative, i.e., $\Delta U_{n\rightarrow n'}=\mathrm{max}[U(n')-U(n), 0]$. (2) Maximal-threshold model: With $U_{M}=\mathrm{max}[U(n)]$, the energy barrier is $\Delta U_{n\rightarrow n'}=U_M-U(n)$. Thus, in this model the protein moves left or right with the same probability. (3)  Intermediate-threshold model: 
Introducing a threshold energy $U_t(\leq U_M)$, the energy barrier is given by $\Delta U_{n\rightarrow n'}=\mathrm{max}[U_M-U_t, U(n')-U(n), 0]$. In the limit of $U_t\to U_M$, this model corresponds to the no-threshold model. (4) The two-regime model: The threshold energy $U_t$ separates the recognition region ($U(n)<U_t$) from the sliding (searching) mode ($U(n)>U_t$). In the former case, a protein can feel a sequence-dependent potential and moves according to the intermediate-threshold model in (3). In the sliding region, a protein performs a sequence-independent diffusion under a flat energy landscape, which is defined as $U(n)=U_{\mathrm{sl}}>U_t$. Practically, $U_\mathrm{sl}$ can be chosen to be $U_M$.

Computational studies of the above four models have shown that the corresponding diffusion dynamics have some common features: In the short-time regime (corresponding to $t<30~\mathrm{ms}$ upon the plug-in of realistic parameter values for the model) the protein diffusion is subdiffusive with the MSD
\begin{equation}
\langle x^2(t) \rangle =2D_\alpha t^\alpha
\end{equation}
with the anomaly exponent $\alpha\sim 0.5\ldots 0.6$. After the cross-over, the protein dynamics asymptotically converges to a Fickian one ($\alpha=1$) in the long-time limit~\cite{wang2011, barbi2004model}. In numerical studies based on realistic parameter values, the behavior of the cross-over time $t_c$ has not been well studied ($t_c$ is about 0.01$\sim$0.1 s). Larger values of $\epsilon/[k_{B}T]$ lead to 
more subdiffusive dynamics, as one can expect, and delay the crossover from subdiffusion to the Fickian dynamics. Interestingly, as the threshold energy $U_{t}$ increases, the anomaly exponent $\alpha$ in the short-time regime gradually converges to unity. This is because the diffusion  rates $r_{n\to n'}$ for all $n$ become similar at the high-threshold energy limit.

The emergence and the characteristics of the subdiffusive motion shown in the short-time limit could be related to the far-from-equilibrium initial condition---which was the uniform distribution---imposed in the simulation. In a theoretical study with similar dynamic models~\cite{saxton2007}, it was shown that the non-equilibrium initial condition significantly affects the transient subdiffusive dynamics. As the distribution of the initial positions of the tracers comes close to the equilibrium distribution, the subdiffusive character in the short-time becomes milder. In any case, a short-time subdiffusion emerges, albeit a mild one, for the models such as the no-threshold model above where the diffusion rate depends on the adjacent energy level~\cite{saxton2007}.


\subsection{Diffusion on a correlated Gaussian energy landscape}
\label{sec:diffcorrelatedlanscape}
The contact range of a protein to a DNA is typically from 5 to 30 bp long~\cite{goychuk2014, barbi2004}. Therefore, when a bound protein moves along the DNA, it has sequence-dependent interactions via the bp binding energy, electrostatic interactions, etc. These DNA-protein interactions can be treated as a spatially disordered potential. With this as a basis, the protein diffusion can be viewed as that of a particle in a one-dimensional spatially correlated random energy landscape. Then in this model, the protein diffusion is described by an overdamped Langevin equation~\cite{goychuk2014}:
\begin{eqnarray}
\label{eqn:randompotential}
\gamma\frac{dx}{dt}=-\frac{\partial U(x)}{\partial x}+\sqrt{2\gamma k_BT}\xi(t),
\end{eqnarray}
where $\gamma$ is the friction coefficient and $\xi(t)$ is the usual white Gaussian noise as defined in Section~III.  Here, the potential energy $U(x)$ is a random potential, with zero mean ($ \overline{U(x)}=0$) and a spatial correlation 
\begin{equation}\label{eq:U}
\overline{ U(x)U(x')  } =\epsilon^{2} g(|x-x'|).
\end{equation}
Here, $g(x)$ is a function describing the decay of the correlation, and $\epsilon$ is the rms of the potential fluctuation (see Eq.~\eqref{eqn:helical}). The overline symbol $\overline{Q}$ indicates the disorder averaging of a physical observable $Q$ over different random potentials, as defined in Section~III. The simplest model for $g$ is the exponentially decaying one $g(x)={\rm exp} (-|x|/\lambda)$ with the correlation length $\lambda$ being the linear size of a DNA--protein contact. The study of this model has shown that the MSD increases as $\overline{\langle x^2(t) \rangle }= 2(k_BT/\gamma) t$ in the short-time limit and asymptotically converges to 
\begin{equation}
\overline{\langle x^2(t) \rangle} \simeq 2(k_BT/\gamma){\rm exp}(-\epsilon^2/[k_BT]^2)t.
\end{equation}
Overall, the MSD is Fickian at both short- and long-time limits, and there is a cross-over non-Fickian regime in the intermediate time. As the ratio of $\epsilon/[k_BT]$ increases, it becomes clearer that the MSD has, transiently, a sub-linear scaling of the form $\overline{\langle x^2(t) \rangle} \sim t^{\alpha(t)}$, where the anomaly exponent is time-dependent $0<\alpha(t)<1$ and $P(x, t)$ becomes non-Gaussian~\cite{goychuk2014, godec2014, banerjee2014}. In the extreme case where $\epsilon/[k_BT]\gg1$, the MSD exhibits Sinai diffusion $\overline{\langle x^2 \rangle} \sim ({\rm ln}~t)^4$ \cite{goychuk2017}.

They also studied the time required for a particle initially localized at the center of a spatial domain $[-\lambda, \lambda]$ to escape from it. They observed that the numerical distribution deviates significantly from the one calculated by averaging the disorder $\psi(t) = \pi\sum_n^\infty (-1)^n (2n+1){\rm exp}[-\pi^2(2n+1)^2 k_BT{\rm e}^{-\epsilon^2/[k_BT]^2}t/(4\gamma\lambda)]$. Indeed, the numerical mean and variance are much smaller and they exhibit a ${\rm exp}(\epsilon/[k_BT])$ dependence rather than the ${\rm exp}(\epsilon^2/[k_BT]^2)$ predicted by averaging the disorder. Thus, such a subdiffusion due to spatial correlations proceeds much faster than one would expect from the renormalization by the disorder.


\subsection{Viscoelastic subdiffusion on a correlated Gaussian energy landscape}
\label{sec:continuousdisorderedsystem}


An extension of the model presented in the previous section VI.~3 was recently made in order to take into account a viscoelastic environment \cite{goychuk2018}. Indeed, the diffusion of the protein can be considered as a viscoelastic subdiffusion of a particle on a random potential $U(x)$~\cite{goychuk2018}, and it satisfies an overdamped generalized Langevin equation \cite{kharchenko2013}



\begin{equation}
\label{eqn:viscorp}
\begin{aligned}
\gamma\frac{dx}{dt}+&\gamma_\alpha\int_0^t {\rm d}\tau \frac{\dot{x}(\tau)}{\Gamma(1-\alpha) (t-\tau)^{\alpha}}= \\
&-\frac{\partial U(x)}{\partial x}+\sqrt{2\gamma k_BT}\xi(t)+\sqrt{\gamma_\alpha k_BT}\xi_\alpha(t)
\end{aligned}
\end{equation}
where the anomaly exponent is in the range of $0<\alpha<1$, and $U(x)$ is the same random potential as in Eqs.~\eqref{eqn:randompotential} and \eqref{eq:U}. Here, the correlation function $g$ was modeled either by an exponential function $g(x)={\rm exp}(-|x|/\lambda)$ or by a power law $g(x)=1/[1+x^2/\lambda^2]^{\gamma/2}$, with the correlation length $\lambda$ typically being the size of a DNA--protein contact. In this equation, the particle diffuses in the random potential $U(x)$ under the thermal noise $\xi(t)$ and the fractional Gaussian noise $\xi_\alpha$ (introduced in Section~III.~2). In order to satisfy the fluctuation--dissipation theorem, the thermal noise $\xi(t)$ is compensated by the damping term given by the memoryless friction coefficient $\gamma$, and the fractional Gaussian noise $\xi_\alpha(t)$ compensated by the memory-existing damping term (the second term on the L.H.S.). 

Eq.~\eqref{eqn:viscorp} with $\alpha=1/2$ was numerically solved, from which the corresponding MSD was studied. Depending on the ratio $\epsilon/[k_BT]$,  the particle exhibits different dynamical behavior. For $\epsilon \approx k_BT$, the MSD is the same as that for the potential-free case: it is initially Fickian and displays a crossover to its asymptotic subdiffusive behavior $\overline{\langle x^2(t)\rangle} =2(k_BT/\gamma_\alpha)t^\alpha/\Gamma(1+\alpha)\sim t^{1/2}$. This suggests that the influence of the random potential is completely negligible. However, when $\epsilon> k_BT$, the MSD displays a time-dependent anomalous behavior, i.e., $\overline{\langle x^2(t)\rangle }\sim t^{\alpha(t)}$: the system starts being diffusive, then enters a transient regime where the exponent $0<\alpha<1$, then eventually, because of the decaying correlations of the potential, the system reaches the subdiffusive regime of the potential-free viscoelastic diffusion $\overline{\langle x^2(t) \rangle} =2(k_BT/\gamma_\alpha)t^\alpha/\Gamma(1+\alpha)$. Thus, on the ensemble level, the viscoelastic subdiffusion of \eqref{eqn:viscorp} is not suppressed by the disorder: this is the opposite of the diffusion on correlated energy landscape introduced in Section~VI.~3 where asymptotically the diffusion is suppressed by the factor ${\rm exp}(-\epsilon^2/(k_BT)^2)$. The effect of the disorder is only visible by the transient regime. Furthermore, this transient regime can last very long, so that the asymptotic subdiffusion is practically never reached. Finally, if  $\epsilon \gg k_BT$ (e.g., $ \epsilon= 10k_BT$), the protein diffusion follows a Sinai subdiffusion $\overline{\langle  x^2(t)\rangle }\sim(\mathrm{ln}~t)^4$ for both functions $g(x)$ with an exponential and a power-law decay.


\section{Summary}

We have provided an overview of the experiments and theoretical models that investigate the one-dimensional diffusion dynamics of DNA-binding proteins from the viewpoint of stochastic processes. The DNA-binding proteins of more than ten different groups have been classified, based on their diffusion properties, as obtained mostly by single-molecule experimental studies.  
The characteristics of their diffusion under various conditions, the experimental techniques employed, and probing limits, were scrutinized in detail. As summarized in Table~1, the 1D motion of a DNA-binding protein exhibits protein-specific and diverse dynamic behaviors. The diffusion of these proteins has been observed to range from a simple Fickian type to non-Fickian anomalous ones that display MSDs characterized by $\langle x^2(t)\rangle \propto t^\alpha$ with $\alpha\neq 1$ or have multiple dynamic states, switching between them during 1D motion, or have time- and sequence-dependent dynamics. Even for the cases with Fickian dynamics, it was found that the 1D motion takes distinct forms, depending on whether it is a simple sliding, rotation-coupled sliding, or hopping. Some proteins have been found to have both sliding and hopping mechanisms, presumably via the conformational change of the protein depending on the ambient conditions, such as the salt and ATP concentrations.


Although unexpected interesting dynamic properties of protein diffusion have been elucidated with the state-of-the-art experimental technique, there are still numerous unexplored or unresolved important questions regarding protein diffusion. For example, the structure of a protein in the so-called interrogating state and the exact mechanism of the interrogation process are not fully understood. Second, a microscopic understanding of the protein diffusion on the DNA is still lacking. It is not clearly understood whether a protein continuously forms and destroys a set of hydrogen bonds while sliding along the DNA and how a protein diffusion is affected by structural changes in the DNA, such as bending, nicking, and bp opening. The effect of the crowded and active \emph{in vivo} environment on the motion of the protein is hardly known. There seems to be a discrepancy between the experiments and the theoretical models. While several microscopic diffusion models (Section~VI) based on a sequence-dependent energy landscape predict a subdiffusive dynamics at a certain timescale, only a few proteins have been experimentally found to exhibit a non-Fickian motion with  $0<\alpha<1$. This could be attributed to the limitation of the spatiotemporal resolution accessed by the current single-particle tracking technique or the gap between the timescales of measurement and the subdiffusive regime, or due to an unknown effect from the DNA--protein interactions (e.g., frequent association/dissociation and the interrogation process). 

Lastly, we note in passing that the experimental study combined with a state-of-the-art statistical analysis~\cite{metzler2014} can provide valuable insight into the mechanisms and stochastic identity of the observed protein diffusion. Apart from the typically examined quantities, such as the MSD and the propagator, one can analyze statistical properties, such as the correlation in the displacements, non-Markovianity, non-ergodicity, and spatiotemporal heterogeneity.  The  corresponding knowledge would help one to set up a realistic model for protein diffusion and to suggest new ideas for experiment. 


\begin{acknowledgments}
This work was supported by the National Research Foundation (NRF) of Korea through NO.~2018R1A6A3A11043366 (O.L), No.~2020R1I1A1A01071790 (X.D), and No.~2017K1A1A2013241 (J-H.J).
\end{acknowledgments}

\bibliography{reference} 
\bibliographystyle{jkps} 
\end{document}